# Unified Formulation of Phase Space Mapping Approaches for Nonadiabatic Quantum Dynamics

*Jian Liu\*, Xin He, Baihua Wu*

Beijing National Laboratory for Molecular Sciences, Institute of Theoretical and Computational Chemistry, College of Chemistry and Molecular Engineering,

Peking University, Beijing 100871, China

\*    Author to whom correspondence should be addressed: jianliupku@pku.edu.cn

**CONSPECTUS**: Nonadiabatic dynamical processes are one of the most important quantum mechanical phenomena in chemical, materials, biological, and environmental molecular systems, where the coupling between different electronic states is either inherent in the molecular structure or induced by the (intense) external field. The curse of dimensionality indicates the intractable exponential scaling of calculation effort with system size and restricts the implementation of "numerically exact" approaches for realistic large systems. The phase space formulation of quantum mechanics offers an important theoretical framework for constructing practical approximate trajectory-based methods for quantum dynamics. This Account reviews our recent progress in phase space mapping theory: a unified framework for constructing the mapping Hamiltonian on phase space for coupled $F$-state systems where the renowned Meyer-Miller Hamiltonian model is a special case, a general phase space formulation of quantum mechanics for nonadiabatic systems where the electronic degrees of freedom are mapped onto constraint space and the nuclear degrees of freedom are mapped onto infinite space, and an isomorphism between the mapping phase space approach for nonadiabatic systems and that for nonequilibrium electron transport processes. While the zero-point-energy parameter is conventionally assumed to be positive, we show that the constraint implied in the conventional Meyer-Miller mapping Hamiltonian requires that such a parameter can be negative as well and lies in $\left(-1/F, +\infty\right)$ for each electronic degree of freedom. More importantly, the zero-point-energy parameter should be interpreted as a special case of a commutator matrix in the comprehensive phase space mapping Hamiltonian for nonadiabatic systems. From the rigorous formulation of mapping phase space, we propose approximate but practical trajectory-based nonadiabatic dynamics methods. The applications to both gas phase and condensed phase problems include the spin-boson model for condensed phase dissipative two-state systems, the three-state photo-dissociation models, the seven-site model of the Fenna-Matthews-Olson monomer in photosynthesis of green sulfur bacteria, the strongly-coupled molecular/atomic matter-optical cavity systems designed for controlling and manipulating chemical dynamical processes, and the Landauer model for a quantum dot state coupled with two electrodes. In these applications the overall performance of our phase space mapping dynamics approach is superior to two prevailing trajectory-based methods, Ehrenfest dynamics and fewest switches surface hopping.

**KEY REFERENCES**

- Liu, J., A Unified Theoretical Framework for Mapping Models for the Multi-State Hamiltonian. *J. Chem. Phys.* **2016**, *145*, 204105.[1] The novel unified framework of the article sets the scene for developing phase space mapping models for the coupled multi-electronic-state Hamiltonian.
- Liu, J., Isomorphism between the Multi-State Hamiltonian and the Second-Quantized Many-Electron Hamiltonian with Only 1-Electron Interactions. *J. Chem. Phys.* **2017**, *146*, 024110.[2] The paper shows that there exists an isomorphism between the mapping phase space approach for nonadiabatic systems and that for nonequilibrium electron transport processes.
- He, X.; Liu, J., A New Perspective for Nonadiabatic Dynamics with Phase Space Mapping Models. *J. Chem. Phys.* **2019**, *151*, 024105.[3] In the article mapping approaches on constraint phase space have first been proposed for nonadiabatic dynamics.

- He, X.; Gong, Z.; Wu, B.; Liu, J., Negative Zero-Point-Energy Parameter in the Meyer-Miller Mapping Model for Nonadiabatic Dynamics. *J. Phys. Chem. Lett.* **2021**, *12*, 2496-2501.[4] The paper presents a new general formulation for constructing exact mapping approaches on constaint phase space for a finite number of discrete electronic states of nonadiabatic systems.

1. **INTRODUCTION**

Because the difference between the mass of an electron and that of a nucleus is at least 3 orders of magnitude, the celebrated Born-Oppenheimer (BO) approximation makes the assumption that the electronic and nuclear motions are separated[5]. In the BO scheme, the adiabatic electronic states are obtained when the coordinates of nuclei are fixed. The potential energy surface (PES) for the relevant adiabatic electronic state is then produced either in advance or on-the-fly as nuclear dynamics is considered.

The BO approximation is, however, not valid in nonadiabatic dynamics that occurs in many important quantum mechanical phenomena, such as photochemistry, electron transport/transfer, and cavity-modified transition processes in chemical, materials, biological, and environmental molecular systems[6-9]. Nonadiabatic dynamics includes quantum mechanical behavior of both electrons and nuclei. In such processes, nuclear dynamics involves two or more coupled electronic states, where the state-state coupling is either inherent in the molecular structure or induced by the (intense) external field. As it is often intractable to use "numerically exact" methods for realistic multi-dimensional anharmonic systems, considerable effort has been focused on developing practical (trajectory-based) methodologies[10-14] to address the fundamental nature of nonadiabatic dynamic processes in complex (large) molecular systems.

The phase space formulation of quantum mechanics[15-17] that employs both coordinate and momentum variables offers a widely useful tool to gain insight on bridging quantum and classical counterpart concepts. Since the renowned work of Meyer and Miller[13] and its successful applications to describing the electronic-to-rotational or electronic-to-vibrational resonance energy transfer in nonadiabatic collision reactions[18, 19], the Meyer-Miller mapping model has offered an important theoretical framework for developing practical trajectory-based nonadiabatic dynamics methods[14, 20-41]. A recent review[28] has briefly summarized the important developments and applications on the Meyer-Miller mapping model.

In this Account, we will report our more recent progress in phase space mapping theory since 2016, which is focused on a novel unified framework for mapping Hamiltonian models[1] which offers a novel way to derive the Meyer-Miller model, a new comprehensive formulation for the one-to-one correspondence mapping onto phase space[1, 3, 4, 42], and an isomorphism between the mapping phase space approach for the coupled multi-state system and that for the second-quantized many-electron Hamiltonian system[2]. In the applications to nonadiabatic transition processes, we will discuss the comparison of the phase space mapping approaches to two prevailing trajectory-based methods, Ehrenfest dynamics[10] and Tully's fewest switches surface hopping (FSSH)[11].

## 2. MEYER−MILLER MAPPING HAMILTONIAN MODEL

In 1927 Dirac demonstrated that the time-dependent Schrödinger equation for an *F*-state quantum system is identical to Hamilton's equations of motion (EOMs) for the action-angle variables[43]. In 1979 Meyer and Miller suggested a heuristic mapping Hamiltonian model with the "Langer correction" for a finite set of electronic states of a molecular system such that both nuclear and electronic degrees of freedom (DOFs) are treated on the same footing for dynamics[13]. They use the diabatic representation for simplicity. The Meyer-Miller Hamiltonian is

$$H_{\text{MM}}(\mathbf{x},\mathbf{p};\mathbf{R},\mathbf{P}) = \frac{1}{2}\mathbf{P}^T\mathbf{M}^{-1}\mathbf{P} + \sum_{n,m=1}^{F}\left[\frac{1}{2}(x^{(n)}x^{(m)} + p^{(n)}p^{(m)}) - \gamma\delta_{nm}\right]V_{nm}(\mathbf{R}) \quad . \quad (1)$$

Here, $\{\mathbf{R}, \mathbf{P}\}$ are the nuclear coordinate and momentum variables (the total number of nuclear DOFs is $N$), $\mathbf{M}$ is the diagonal 'mass matrix' with elements $\{m_j\}$, $\{\mathbf{x}, \mathbf{p}\} = \{x^{(1)}, \cdots, x^{(F)}, p^{(1)}, \cdots, p^{(F)}\}$ are the mapping coordinate and momentum variables for the finite set of electronic states (with $F$ the total number of states), and parameter $\gamma = 1/2$ in Meyer and Miller's original version where the Langer correction [for the zero-point-energy (ZPE)] is employed[13]. Equation 1 is the mapping Hamiltonian for the coupled $F$-electronic-state Hamiltonian in quantum mechanics

$$\hat{H} = \sum_{n,m=1}^{F} H_{nm}(\hat{\mathbf{R}}, \hat{\mathbf{P}})|n\rangle\langle m| = \sum_{n,m=1}^{F}\left[\frac{1}{2}\hat{\mathbf{P}}^T \mathbf{M}^{-1}\hat{\mathbf{P}}\delta_{nm} + V_{nm}(\hat{\mathbf{R}})\right]|n\rangle\langle m| \quad , \qquad (2)$$

where the $F$ electronic states form an orthogonal complete basis set, that is,

$$\langle m|n\rangle = \delta_{mn}, \quad \hat{I}_{ele} = \sum_{n=1}^{F}|n\rangle\langle n| \quad . \qquad (3)$$

Here $\hat{I}_{ele}$ is the identity operator of the electronic state space, and $\{V_{nm}(\mathbf{R}) = V_{mn}(\mathbf{R})\}$ are the elements of the real symmetric matrix for the potential energy operator.

In Ref. [20] Stock and Thoss used the oscillator model of angular momentum proposed by Schwinger and suggested the mapping relations

$$|n\rangle \mapsto \underbrace{|0_1 \cdots 1_n \cdots 0_F\rangle}_{F-\text{states}} \qquad (4)$$

and

$$|n\rangle\langle m| \mapsto \hat{a}_n^+ \hat{a}_m \quad , \qquad (5)$$

where conventional commutation relations of the harmonic-oscillator creation and annihilation operators $\{\hat{a}_n^+, \hat{a}_n\}$ are

$$\left[\hat{a}_m, \hat{a}_n^+\right] = \delta_{mn} \quad (\forall m, n) \quad . \tag{6}$$

The Hamiltonian operator of eq 2 is then equivalent to

$$\hat{H} = \sum_{n,m=1}^{F} H_{nm}\left(\hat{\mathbf{R}}, \hat{\mathbf{P}}\right) \hat{a}_n^+ \hat{a}_m \quad . \tag{7}$$

After the creation/annihilation operators are transformed into pairs of operators

$$\begin{aligned}\hat{x}_n &= \frac{\hat{a}_n + \hat{a}_n^+}{\sqrt{2}} \\ \hat{p}_n &= \frac{\hat{a}_n - \hat{a}_n^+}{\sqrt{2}i}\end{aligned} \quad , \tag{8}$$

the coupled $F$-electronic-state Hamiltonian operator eq 7 becomes the Meyer-Miller mapping Hamiltonian eq 1. Equation 6 yields

$$\left[\hat{x}_m, \hat{p}_n\right] = i\delta_{mn} \quad (\forall m, n) \quad , \tag{9}$$

where $\delta_{mn}$ is the Kronecker delta. Substitution of eqs 8 and 9 into eq 7 leads to eq 1, which demonstrates that parameter $\gamma = 1/2$ comes from the commutation relation eq 9 and is the ZPE of the harmonic oscillator for each underlying electronic DOF[20]. It is then evident that eq 1 is an exact mapping Hamiltonian model for eq 2, the coupled $F$-electronic-state Hamiltonian in quantum mechanics. In practice, the ZPE parameter is chosen to be $1/3$, $(\sqrt{3}-1)/2$, $(\sqrt{F+1}-1)/F$, and other non-negative values in its semiclassical/quasiclassical applications[20, 21, 25-32, 37, 44].

Because there exist an infinite number of energy eigenstates of a harmonic oscillator, the mapping of eqs 4 and 5 is then restricted onto the oscillator subspace with a single excitation, that is, only ground and first excited levels are employed. Evaluation of physical properties or time correlation functions invokes a space of singly excited oscillators (SEOs) [14, 20-23, 31, 32, 34, 37, [38, 45, 46]. Appendix A of Ref. [1] evidently shows that eq 6 or eq 9 holds only when an infinite number of eigenstates of a harmonic oscillator are involved. In Ref. [26] Cotton and Miller pointed out that the Meyer-Miller mapping model "is not the most natural one". They suggested a spin mapping model (SPM)[26], but SPM is not exact for general multi-electronic-state systems even in the frozen nuclei limit (i.e., variables of nuclear DOFs are fixed.)[1]. (See Figure 1.) This explains why SPM performs worse than the Meyer-Miller model. It is evidently nontrivial to obtain exact mapping Hamiltonian models on phase space for the coupled multi-state Hamiltonian. To the best of our knowledge, except the Meyer-Miller model[13, 20], no exact phase space mapping Hamiltonian models for nonadiabatic systems were explicitly proposed before 2016.

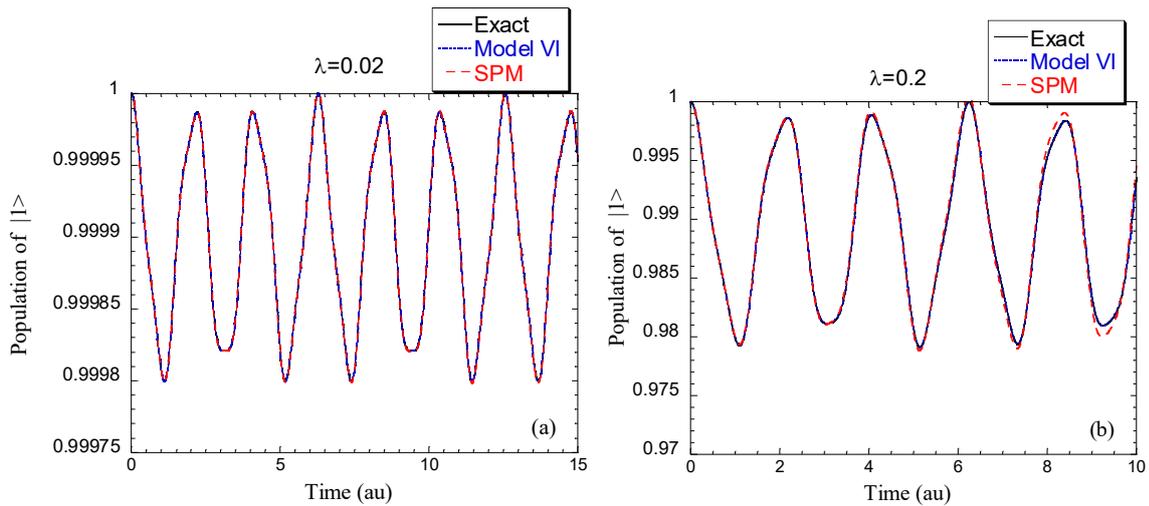

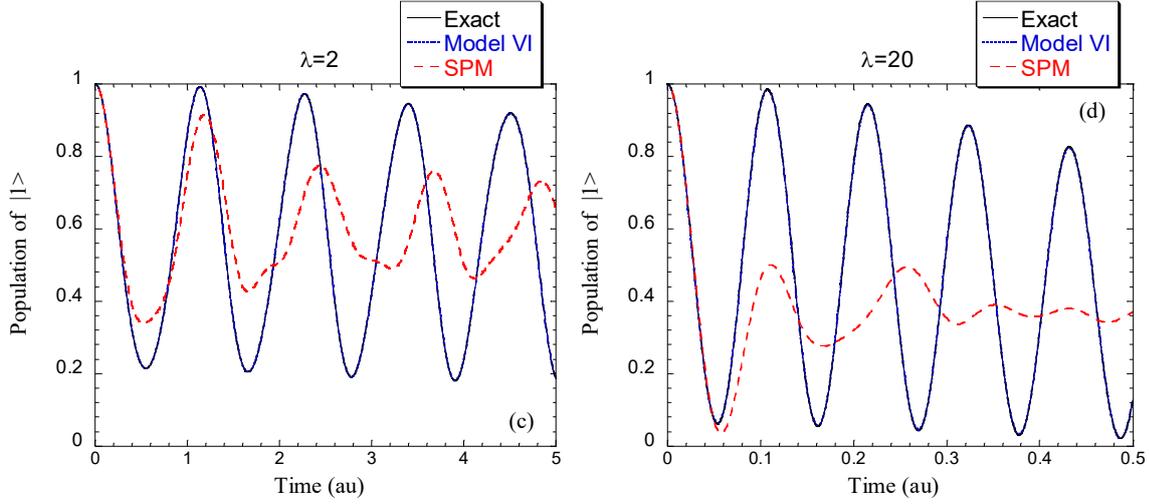

**Figure 1** (Color). Comparison between an exact mapping model of Ref. [1] and SPM of Ref. [26] for population dynamics of state $|1\rangle$ for a 3-state Hamiltonian system[1]. As the state-state coupling $\lambda$ increases, the performance of SPM becomes worse. (Reproduced with permission from Ref. [1]. Copyright 2016 American Institute of Physics.)

## 3. UNIFIED FORMULATION FOR PHASE SPACE MAPPING APPROACHES

### 3.1 Unified Framework for Mapping Hamiltonian Models

In Ref. [1] we proposed a unified framework for constructing exact phase space mapping Hamiltonian models. The three key elements are:

1) We first introduce a vacuum (or reference) state $|\bar{0}\rangle \mapsto \underbrace{|0_1 \cdots 0_n \cdots 0_F\rangle}_{F-\text{states}}$ such that state $n$ of eq 4 is represented by a single excitation from the vacuum state, that is,

$$|n\rangle = \hat{a}_n^+ |\bar{0}\rangle \quad . \qquad (10)$$

Here, an excitation stands for the occupation of the corresponding state, and the vacuum state, $|\bar{0}\rangle$, is orthogonal to any occupied state $|n\rangle$.

2) Because only a single excitation is invoked, we can define the creation and annihilation operators as

$$\begin{aligned}\hat{a}_n^+ &= |n\rangle\langle\bar{0}| \\ \hat{a}_n &= |\bar{0}\rangle\langle n|\end{aligned} \quad . \tag{11}$$

Equation 11 leads to eq 5 as well as eq 7. It is straightforward to derive the commutation and anti-commutation relations from eq 11, which demonstrate that the underlying DOFs are neither bosons nor fermions[1]. Define

$$\begin{aligned}\hat{\sigma}_x^{(n)} &= \hat{a}_n + \hat{a}_n^+ \\ \hat{\sigma}_y^{(n)} &= \frac{\hat{a}_n - \hat{a}_n^+}{i}\end{aligned} \quad . \tag{12}$$

$\{\hat{\sigma}_x^{(n)}, \hat{\sigma}_y^{(n)}\}$ stand for the Pauli matrices (for a spin 1/2 particle) in the x, y directions, respectively.

3) We then obtain equivalent expressions of the coupled multi-state Hamiltonian operator of eq 7 in terms of only $\{\hat{\sigma}_x^{(n)}, \hat{\sigma}_y^{(n)}\}$ and develop the criteria for obtaining exact phase space mapping models.

The novel framework involves *only* quantum operators for constructing exact mapping Hamiltonian models. In contrast, the conventional way in the literature always involves a space of SEOs for implementing the Meyer-Miller mapping model[20, 21].

An equivalent representation of the Hamiltonian operator of eq 7 reads

$$\hat{H} = \sum_{n,m=1}^{F} H_{nm}\left(\hat{\mathbf{R}}, \hat{\mathbf{P}}\right) \left\{ \frac{1}{4}\left(\hat{\sigma}_x^{(n)} \hat{\sigma}_x^{(m)} + \hat{\sigma}_y^{(n)} \hat{\sigma}_y^{(m)}\right) - \frac{i}{4}\left(\hat{\sigma}_y^{(n)} \hat{\sigma}_x^{(m)} - \hat{\sigma}_x^{(n)} \hat{\sigma}_y^{(m)}\right) \right\} . \quad (13)$$

A change of variables

$$x^{(n)} = \frac{\sigma_x^{(n)}}{\sqrt{2}}, \quad p^{(n)} = \frac{\sigma_y^{(n)}}{\sqrt{2}} \quad (14)$$

in eq 13 yields

$$H(\mathbf{x},\mathbf{p};\boldsymbol{\Gamma};\mathbf{R},\mathbf{P}) = \sum_{n,m=1}^{F} \left[ \frac{1}{2}\left(x^{(n)} x^{(m)} + p^{(n)} p^{(m)}\right) - \Gamma_{nm} \right] H_{nm}(\mathbf{R},\mathbf{P}) . \quad (15)$$

Here

$$\Gamma_{nm} = \frac{i}{4}\left(\hat{\sigma}_y^{(n)} \hat{\sigma}_x^{(m)} - \hat{\sigma}_x^{(n)} \hat{\sigma}_y^{(m)}\right) \quad (16)$$

is the element in the $n$th row and $m$th column of commutator matrix $\boldsymbol{\Gamma}$. Equation 16 implies

$$\hat{x}^{(n)} \hat{p}^{(m)} - \hat{p}^{(n)} \hat{x}^{(m)} = 2i\Gamma_{nm} , \quad (17)$$

or the commutation relation

$$\left[\hat{x}^{(n)}, \hat{p}^{(m)}\right] + \left[\hat{x}^{(m)}, \hat{p}^{(n)}\right] = 2i\left(\Gamma_{nm} + \Gamma_{mn}\right) . \quad (18)$$

Equation 18 is the commutation relation between the position and momentum operators of the quasi-particles of the electronic mapping DOFs. It is evident that the conventional canonical commutation relation eq 9 is only a special case of eq 18.

When the equality

$$\sum_{n=1}^{F} \left[ \frac{\left(x^{(n)}\right)^2 + \left(p^{(n)}\right)^2}{2} - \Gamma_{nn} \right] = 1 \quad (19)$$

holds for the mapping phase variables of the electronic DOFs, the mapping Hamiltonian becomes

$$H(\mathbf{x},\mathbf{p};\mathbf{\Gamma};\mathbf{R},\mathbf{P}) = \frac{1}{2}\mathbf{P}^T\mathbf{M}^{-1}\mathbf{P} + \sum_{n,m=1}^{F}\left[\frac{1}{2}(x^{(n)}x^{(m)} + p^{(n)}p^{(m)}) - \Gamma_{nm}\right]V_{mn}(\mathbf{R}) \ . \quad (20)$$

The conventional Meyer-Miller mapping Hamiltonian eq 1 is intrinsically a special case with $\Gamma_{nm} = \gamma\delta_{nm}$ for eq 20, the general and comprehensive mapping Hamiltonian[1, 3, 42] (i.e., Model II of Ref. [1]).

The unified framework of Ref. [1] provides the scene for obtaining more equivalent representations of the $F$-state quantum system and their corresponding exact phase space mapping Hamiltonian models. As shown in Ref. [1], the mapping in analogy to angular momentum produces leads to Model I,

$$H(\mathbf{x},\mathbf{p_x},\mathbf{y},\mathbf{p_y};\mathbf{\Gamma};\mathbf{R},\mathbf{P}) = \sum_{n,m=1}^{F}\left[\left(x^{(n)}p_y^{(m)} - y^{(n)}p_x^{(m)}\right) - \Gamma_{nm}\right]H_{nm}(\mathbf{R},\mathbf{P}) \ . \quad (21)$$

Models III, IV, V and VI for the coupled multi-state Hamiltonian operator (proposed in Ref. [1]) are

$$\begin{aligned}H = &\sum_{n=1}^{F}\left(\frac{\left(x^{(n)}+p_y^{(n)}\right)^2 + \left(y^{(n)}-p_x^{(n)}\right)^2}{4} - \Gamma_{nn}\right)H_{nn}(\mathbf{R},\mathbf{P}) \\ &+ \sum_{n\neq m}\left(x^{(n)}p_y^{(m)} - y^{(m)}p_x^{(n)} - \Gamma_{nm}\right)H_{nm}(\mathbf{R},\mathbf{P})\end{aligned} \quad (22)$$

$$\begin{aligned}H = &\sum_{n=1}^{F}\left(\frac{\left(x^{(n)}+p_y^{(n)}\right)^2 + \left(y^{(n)}-p_x^{(n)}\right)^2}{4} - \Gamma_{nn}\right)H_{nn}(\mathbf{R},\mathbf{P}) \\ &+ \sum_{n<m}\left(\left(x^{(n)}x^{(m)} + p_x^{(n)}p_x^{(m)}\right) + \left(y^{(n)}y^{(m)} + p_y^{(n)}p_y^{(m)}\right) - \Gamma_{nm} - \Gamma_{mn}\right)H_{nm}(\mathbf{R},\mathbf{P})\end{aligned} \quad (23)$$

$$\begin{aligned}H = &\sum_{n=1}^{F}\left(\frac{\left(x^{(n)}+p_y^{(n)}\right)^2 + \left(y^{(n)}-p_x^{(n)}\right)^2}{4} - \Gamma_{nn}\right)H_{nn}(\mathbf{R},\mathbf{P}) \\ &+ \sum_{n<m}\left(\frac{\left(x^{(n)}+p_y^{(n)}\right)\left(x^{(m)}+p_y^{(m)}\right) + \left(y^{(n)}-p_x^{(n)}\right)\left(y^{(m)}-p_x^{(m)}\right)}{2} - \Gamma_{nm} - \Gamma_{mn}\right)H_{nm}(\mathbf{R},\mathbf{P})\end{aligned} \quad (24)$$

and

$$H = \sum_{n=1}^{F}\left(x^{(n)}p_y^{(n)} - y^{(n)}p_x^{(n)} - \Gamma_{nn}\right)H_{nn}(\mathbf{R},\mathbf{P})$$
$$+ \sum_{n<m}\left(\left(x^{(n)}x^{(m)} + p_x^{(n)}p_x^{(m)}\right) + \left(y^{(n)}y^{(m)} + p_y^{(n)}p_y^{(m)}\right) - \Gamma_{nm} - \Gamma_{mn}\right)H_{nm}(\mathbf{R},\mathbf{P}) \quad , \quad (25)$$

respectively. Similar to eq 19, when we have the equality for the mapping phase variables of the electronic DOFs

$$\sum_{n=1}^{F}\left(x^{(n)}p_y^{(n)} - y^{(n)}p_x^{(n)} - \Gamma_{nn}\right) = 1 \quad (26)$$

for eq 21 or eq 25 or

$$\sum_{n=1}^{F}\frac{\left(x^{(n)} + p_y^{(n)}\right)^2 + \left(y^{(n)} - p_x^{(n)}\right)^2}{4} - \Gamma_{nn} = 1 \quad (27)$$

for eq 22, 23, or 24, the term related to kinetic energy $\mathbf{P}^T\mathbf{M}^{-1}\mathbf{P}/2$, of the mapping Hamiltonian does not include any electronic state mapping variables[3]. For example, the mapping Hamiltonian eq 21 becomes

$$H(\mathbf{x},\mathbf{p_x},\mathbf{y},\mathbf{p_y};\Gamma;\mathbf{R},\mathbf{P}) = \frac{1}{2}\mathbf{P}^T\mathbf{M}^{-1}\mathbf{P} + \sum_{n,m=1}^{F}\left[(x^{(n)}p_y^{(m)} - y^{(n)}p_x^{(m)}) - \Gamma_{nm}\right]V_{mn}(\mathbf{R}) \quad . (28)$$

Here we demonstrate two more mapping Hamiltonian models. Consider an equivalent representation of the Hamiltonian operator of eq 7,

$$\hat{H} = \sum_{n}\frac{1}{4}\left(\left[\hat{\sigma}_x^{(n)},\hat{\sigma}_x^{(n)}\right]_+ + \left[\hat{\sigma}_y^{(n)},\hat{\sigma}_y^{(n)}\right]_+\right)H_{nn}(\hat{\mathbf{R}},\hat{\mathbf{P}})$$
$$+ \sum_{n<m}\frac{1}{2}\left(\left[\hat{\sigma}_x^{(n)},\hat{\sigma}_x^{(m)}\right]_+ + \left[\hat{\sigma}_y^{(n)},\hat{\sigma}_y^{(m)}\right]_+\right)H_{nm}(\hat{\mathbf{R}},\hat{\mathbf{P}}) \quad . \quad (29)$$

Employing the mapping strategy by analogy with the classical vector as described in Section III-C of Ref. [1] leads to

$$H(\mathbf{x},\mathbf{p}_x,\mathbf{y},\mathbf{p}_y;\boldsymbol{\Gamma};\mathbf{R},\mathbf{P}) = \sum_{n,m=1}^{F}\left[\frac{\left(x^{(n)}x^{(m)}+p_x^{(n)}p_x^{(m)}\right)+\left(y^{(n)}y^{(m)}+p_y^{(n)}p_y^{(m)}\right)}{2}-\Gamma_{nm}\right]H_{nm}(\mathbf{R},\mathbf{P}) \quad . \quad (30)$$

Making the transformation of variables[1]

$$\mathbf{x}=\mathbf{q},\ \mathbf{y}=-\mathbf{p}_\mathbf{q},\ \mathbf{p}_\mathbf{x}=\mathbf{p}_\mathbf{r},\ \mathbf{p}_\mathbf{y}=\mathbf{r} \qquad (31)$$

in eq 21 yields

$$H(\mathbf{q},\mathbf{p}_\mathbf{q},\mathbf{r},\mathbf{p}_\mathbf{r};\boldsymbol{\Gamma};\mathbf{R},\mathbf{P}) = \sum_{n,m=1}^{F}\left[\left(q^{(n)}r^{(m)}+p_q^{(n)}p_r^{(m)}\right)-\Gamma_{nm}\right]H_{nm}(\mathbf{R},\mathbf{P}) \quad . \quad (32)$$

Switching between $\mathbf{q}$ and $\mathbf{r}$ and that between $\mathbf{p}_\mathbf{q}$ and $\mathbf{p}_\mathbf{r}$ make no difference. It is then trivial to obtain from eq 32

$$H(\mathbf{q},\mathbf{p}_\mathbf{q},\mathbf{r},\mathbf{p}_\mathbf{r};\boldsymbol{\Gamma};\mathbf{R},\mathbf{P}) = \sum_{n,m=1}^{F}\left[\frac{\left(q^{(n)}r^{(m)}+p_q^{(n)}p_r^{(m)}\right)+\left(r^{(n)}q^{(m)}+p_r^{(n)}p_q^{(m)}\right)}{2}-\Gamma_{nm}\right]H_{nm}(\mathbf{R},\mathbf{P}) \quad . \quad (33)$$

It is evident that the two mapping Hamiltonian models proposed in Ref. [47] are simply a special case of eq 30 and of eq 33, which are yielded in the unified framework for mapping Hamiltonian models for coupled multi-state systems[1]. The Clifford algebra can be used for the mapping Hamiltonian models (eqs 21-25 and eqs 32-33) that involve $4F$ mapping phase variables for the electronic DOFs. When different mapping Hamiltonian models in the unified framework are treated in the same fashion, they in principle produce the same results, regardless of different convergence performance.

**3.2  General Formulation of the One-to-One Correspondence Mapping on Phase Space for Systems Involving Both Continuous and Discrete DOFs**

In addition to the mapping Hamiltonian, evaluation of physical properties lies in the center of the phase space formulation of quantum mechanics. As first pointed out for general coupled multi-state systems in Ref. [3], eq 19, 26, or 27 suggests the constraint phase space for the mapping variables. This is very different from the classification scheme[16, 17] of the conventional mapping onto full phase space with infinite boundaries[15, 16] for quantum systems that involve the continuous coordinate space. In Refs. [3, 4] we have first proposed a unified formulation for the one-to-one correspondence mapping on phase space, which is capable of treating quantum systems represented in the finite-dimensional as well as infinite-dimensional Hilbert space[1, 3, 4, 42]. The general phase space formulation provides a useful tool for studying nonadiabatic dynamics, where a finite set of electronic states as well as continuous nuclear DOFs is involved.

For demonstration we use the mapping approach eq 20 that is reminiscent of the Meyer-Miller model. Consider the simplest case $\Gamma_{nm} = \delta_{nm}\gamma$ ($\forall n, m$) for commutator matrix $\boldsymbol{\Gamma}$. Because the electronic state mapping variables should satisfy eq 19, a simple strategy is to employ the constraint space,

$$\mathcal{S}(\mathbf{x}, \mathbf{p}) : \delta\left(\sum_{n=1}^{F} \frac{\left(x^{(n)}\right)^2 + \left(p^{(n)}\right)^2}{2} - (1 + F\gamma)\right) , \quad (34)$$

for developing the formulation for evaluation of physical observables. In eq 34, the possible value of parameter $\gamma$ lies in $\left(-\frac{1}{F}, \infty\right)$. The trace of a product of two operators is expressed on phase space as

$$\mathrm{Tr}_{n,e}\left[\hat{A}\hat{B}\right] = \int (2\pi\hbar)^{-N} \mathrm{d}\mathbf{R} \mathrm{d}\mathbf{P} \int_{\mathcal{S}(\mathbf{x},\mathbf{p})} F \mathrm{d}\mathbf{x} \mathrm{d}\mathbf{p} A(\mathbf{R}, \mathbf{P}; \mathbf{x}, \mathbf{p}) \tilde{B}(\mathbf{R}, \mathbf{P}; \mathbf{x}, \mathbf{p}) \quad (35)$$

where

$$A(\mathbf{R}, \mathbf{P}; \mathbf{x}, \mathbf{p}) = \mathrm{Tr}_{n,e}\left[\hat{A}\hat{K}_{nuc}(\mathbf{R}, \mathbf{P}) \otimes \hat{K}_{ele}(\mathbf{x}, \mathbf{p})\right] , \quad (36)$$

$$\tilde{B}(\mathbf{R},\mathbf{P};\mathbf{x},\mathbf{p})=\mathrm{Tr}_{n,e}\left[\hat{K}_{nuc}^{-1}(\mathbf{R},\mathbf{P})\otimes\hat{K}_{ele}^{-1}(\mathbf{x},\mathbf{p})\hat{B}\right], \qquad (37)$$

$(2\pi\hbar)^{-N}d\mathbf{R}d\mathbf{P}\otimes F d\mathbf{x}d\mathbf{p}$ represents the invariant measure on the mapping phase space for the nuclear and electronic DOFs, $\mathrm{Tr}_n$ stands for the trace over the nuclear DOFs, and $\mathrm{Tr}_e$ is the trace over the $F$ electronic states. The inverse one-to-one correspondence mapping from phase space function $A(\mathbf{R},\mathbf{P};\mathbf{x},\mathbf{p})$ (or $\tilde{B}(\mathbf{R},\mathbf{P};\mathbf{x},\mathbf{p})$) of eq 36 to operator $\hat{A}$ (or $\hat{B}$) is

$$\begin{aligned}\hat{A} &= \int (2\pi\hbar)^{-N}\mathrm{d}\mathbf{R}\mathrm{d}\mathbf{P}\int_{\mathcal{S}(\mathbf{x},\mathbf{p})} F\mathrm{d}\mathbf{x}\mathrm{d}\mathbf{p}\, A(\mathbf{R},\mathbf{P};\mathbf{x},\mathbf{p})\hat{K}_{nuc}^{-1}(\mathbf{R},\mathbf{P})\otimes\hat{K}_{ele}^{-1}(\mathbf{x},\mathbf{p})\\ \hat{B} &= \int (2\pi\hbar)^{-N}\mathrm{d}\mathbf{R}\mathrm{d}\mathbf{P}\int_{\mathcal{S}(\mathbf{x},\mathbf{p})} F\mathrm{d}\mathbf{x}\mathrm{d}\mathbf{p}\, \tilde{B}(\mathbf{R},\mathbf{P};\mathbf{x},\mathbf{p})\hat{K}_{nuc}(\mathbf{R},\mathbf{P})\otimes\hat{K}_{ele}(\mathbf{x},\mathbf{p})\end{aligned}. \qquad (38)$$

The integral over constraint space $\mathcal{S}(\mathbf{x},\mathbf{p})$ reads

$$\int_{\mathcal{S}(\mathbf{x},\mathbf{p})} F\mathrm{d}\mathbf{x}\mathrm{d}\mathbf{p}\, g(\mathbf{x},\mathbf{p}) = \frac{\int F\mathrm{d}\mathbf{x}\mathrm{d}\mathbf{p}\,\delta\left(\sum_{n=1}^{F}\left[\frac{(x^{(n)})^2+(p^{(n)})^2}{2}\right]-\left(1+\sum_{n=1}^{F}\Gamma_{nn}\right)\right)g(\mathbf{x},\mathbf{p})}{\int \mathrm{d}\mathbf{x}\mathrm{d}\mathbf{p}\,\delta\left(\sum_{n=1}^{F}\left[\frac{(x^{(n)})^2+(p^{(n)})^2}{2}\right]-\left(1+\sum_{n=1}^{F}\Gamma_{nn}\right)\right)}. \qquad (39)$$

Because the nuclear DOFs involve infinite energy levels, their integrals are over the mapping nuclear phase space with infinite boundaries. The classification scheme[16, 17] can be recast into the definition of the mapping kernel for the nuclear DOFs (in eq 36)

$$\hat{K}_{nuc}(\mathbf{R},\mathbf{P}) = \left(\frac{\hbar}{2\pi}\right)^{N}\int\mathrm{d}\boldsymbol{\zeta}\int\mathrm{d}\boldsymbol{\eta}\, e^{i\boldsymbol{\zeta}\cdot(\hat{\mathbf{R}}-\mathbf{R})+i\boldsymbol{\eta}\cdot(\hat{\mathbf{P}}-\mathbf{P})}f(\boldsymbol{\zeta},\boldsymbol{\eta}), \qquad (40)$$

and that of the inverse kernel (in eq 37)

$$\hat{K}_{nuc}^{-1}(\mathbf{R},\mathbf{P}) = \left(\frac{\hbar}{2\pi}\right)^{N}\int\mathrm{d}\boldsymbol{\zeta}\int\mathrm{d}\boldsymbol{\eta}\, e^{i\boldsymbol{\zeta}\cdot(\hat{\mathbf{R}}-\mathbf{R})+i\boldsymbol{\eta}\cdot(\hat{\mathbf{P}}-\mathbf{P})}\left[f(-\boldsymbol{\zeta},-\boldsymbol{\eta})\right]^{-1}. \qquad (41)$$

In eq 40 and eq 41 $f(\boldsymbol{\zeta},\boldsymbol{\eta})$ is a scalar function that defines the mapping nuclear phase space of choice. For example, the Wigner function[15] takes

$$f(\zeta, \eta) = 1 \quad . \tag{42}$$

In eqs 35, 38, 40 and 41, the integrals for the mapping phase variables for the finite set of electronic states are over constraint space $\mathcal{S}(\mathbf{x}, \mathbf{p})$. As derived first in Appendix A of Ref. [3] and then in the Supporting Information of Ref. [4], the mapping kernel for the $F$ electronic states (in eq 36) is

$$\hat{K}_{ele}(\mathbf{x}, \mathbf{p}) = \sum_{n,m=1}^{F} \left[ \frac{1}{2}\left(x^{(n)} + ip^{(n)}\right)\left(x^{(m)} - ip^{(m)}\right) - \gamma \delta_{nm} \right] |n\rangle\langle m| \tag{43}$$

and the corresponding inverse kernel (in eq 37) is

$$\hat{K}_{ele}^{-1}(\mathbf{x}, \mathbf{p}) = \sum_{n,m=1}^{F} \left[ \frac{1+F}{2(1+F\gamma)^2}\left(x^{(n)} + ip^{(n)}\right)\left(x^{(m)} - ip^{(m)}\right) - \frac{1-\gamma}{1+F\gamma} \delta_{nm} \right] |n\rangle\langle m| \quad . \tag{44}$$

As pointed out in Ref. [42], it is trivial to show that the Q-version, W-version, or P-version of Ref. [36] corresponds to parameter $\gamma = 0, \left(\sqrt{F+1}-1\right)/F$, or 1 of the exact phase space mapping formulation that we *first* proposed in Refs. [1, 3] and then in Ref. [4], respectively. It will be interesting to use our general phase space mapping formulation to include/reformulate other approaches that use the Meyer-Miller mapping model[21, 25, 28-30, 34, 37, 41, 45, 48].

Equation 35 also sets the scene for expressing real time dynamics in the phase space formulation of quantum mechanics. An exact expression of the time correlation function of nonadiabatic systems

$$C_{AB}(t) = \text{Tr}_{n,e}\left[\hat{A}(0)\hat{B}(t)\right] \tag{45}$$

is

$$C_{AB}(t) = \int (2\pi\hbar)^{-N} d\mathbf{R}d\mathbf{P} \int_{\mathcal{S}(\mathbf{x},\mathbf{p})} F d\mathbf{x}d\mathbf{p} A(\mathbf{R}, \mathbf{P}; \mathbf{x}, \mathbf{p}) \tilde{B}(\mathbf{R}, \mathbf{P}; \mathbf{x}, \mathbf{p}; t) \quad , \tag{46}$$

where

$$\tilde{B}(\mathbf{R},\mathbf{P};\mathbf{x},\mathbf{p};t) = \mathrm{Tr}_{n,e}\left[\hat{K}_{nuc}^{-1}(\mathbf{R},\mathbf{P})\otimes\hat{K}_{ele}^{-1}(\mathbf{x},\mathbf{p})\hat{B}(t)\right] \ . \tag{47}$$

In eq 45 the Heisenberg operator $\hat{B}(t) = e^{i\hat{H}t/\hbar}\hat{B}e^{-i\hat{H}t/\hbar}$ is used. It is straightforward to express the quantum Liouville equation for $\tilde{B}(\mathbf{R},\mathbf{P};\mathbf{x},\mathbf{p};t)$ of eq 47 in the general phase space formulation. As long as we exactly solve the EOMs (for nuclear and electronic DOFs) in eq 47, the formulation of the correlation function eq 46 is exact for describing nonadiabatic systems[3, 4].

It is, however, often challenging if not at all impossible to exactly solve the EOMs when both nuclear and electronic DOFs are coupled. When we make the trajectory-based dynamics approximation, eq 46 is then recast into

$$C_{AB}(t) = \int (2\pi\hbar)^{-N}\mathrm{d}\mathbf{R}\mathrm{d}\mathbf{P}\int_{\mathcal{S}(\mathbf{x},\mathbf{p})}F\mathrm{d}\mathbf{x}\mathrm{d}\mathbf{p}A(\mathbf{R},\mathbf{P};\mathbf{x},\mathbf{p})\tilde{B}(\mathbf{R}_t,\mathbf{P}_t;\mathbf{x}_t,\mathbf{p}_t) \ . \tag{48}$$

When only the electronic state variables evolve with time, that is, in the frozen nuclei limit, trajectory-based dynamics governed by Hamilton's EOMs of the mapping Hamiltonian (eq 1, eq 15, and other exact mapping Hamiltonian models) produce exact results, that is, eq 48 is equivalent to eq 46. When both nuclear and electronic DOFs are involved, the independent trajectory generated by the mapping Hamiltonian of Section 3.1 is an approximation to the quantum Liouville equation of the corresponding phase space, that is, eq 48 is an approximation to eq 46. The formulation of the correlation function eq 48 is often expressed on constraint space $\mathcal{S}(\mathbf{x},\mathbf{p})$ and Wigner phase space for electronic and nuclear DOFs, respectively. The extended classical mapping model (eCMM) approach[3, 4] utilizes the Meyer-Miller Hamiltonian eq 1 to yield the EOMs of the trajectory. In contrast, the eCMM with commutator variables (eCMMcv) employs the comprehensive mapping Hamiltonian eq 20 to produce trajectory-based dynamics, where initial values for commutator variables are given by

$$\tilde{\mathcal{S}}(\mathbf{x},\mathbf{p};\Gamma) : \delta\left(\sum_{n=1}^{F} \frac{(x^{(n)})^2 + (p^{(n)})^2}{2} - (1+F\gamma)\right)$$
$$\times \prod_{n=1}^{F}\left(\delta\left(\Gamma_{nn} + \delta_{n,j_{occ}} - \frac{(x^{(n)})^2 + (p^{(n)})^2}{2}\right)\prod_{k\neq n}^{F}\delta(\Gamma_{nk})\right). \quad (49)$$

Here $j_{occ}$ denotes the index of the initially occupied state. Both the frozen-nuclei limit and Born-Oppenheimer limit are satisfied in the eCMMcv approach. Our general phase space formulation is not limited to the mapping of a finite set of states onto constraint phase space eq 34 or eq 49. Other options that satisfy eq 19, eq 26, or eq 27 are possible for the constraint space, upon which the one-to-one correspondence mapping can be established. More discussion on this will be available in a forthcoming paper.

Finally, we note that it is easy to extend the phase space mapping approach to the adiabatic representation or other representations. As shown in the Supporting Information of Ref. [42], we can directly apply the strategy of Ref. [29] to the comprehensive mapping Hamiltonian model (of Section 3.1) in the adiabatic representation. The EOMs of the trajectory in the phase space mapping approach are then independent of the electronic representation of choice.

## 4. APPLICATIONS TO NONADIABATIC SYSTEMS

The preceding discussion has reviewed the unified framework for phase space mapping approaches for nonadiabatic quantum dynamics. Below we highlight a few illustrative applications of the eCMM and eCMMcv approaches. Ehrenfest dynamics[10] or FSSH[11] is also implemented for comparison. The FSSH results are obtained first in the adiabatic representation, then projected to the diabatic representation.

### 4.1 The Spin-Boson Model

The spin-boson model is a prototype model for such as electron transfer/transport processes. It depicts a two-electronic-state system coupled with a harmonic vibrational bath environment.

Such a type of model involves key features of nonadiabatic quantum systems in condensed phase[49]. As "numerically exact" results are often available[50-52], it offers a benchmark model for testing the performance of trajectory-based nonadiabatic dynamics methods. Fully converged results are obtained when 300 bath modes (i.e., nuclear DOFs) are employed for the benchmark tests of the spin-boson model[4]. More details on the parameters of the spin-boson model and on implementation of eCMM are presented in Ref. [3] and Ref. [4]. Figure 2 demonstrates that the six mapping Hamiltonian models proposed in Ref. [1] produce the same converged results when they are treated in the same fashion. The conclusion is true for any other mapping Hamiltonian models included in the unified framework[1] as discussed in Section 3.1.

In all conventional approaches where the Meyer-Miller Hamiltonian eq 1 was proposed or derived[13, 20, 21], parameter $\gamma$ of eq 1 was viewed as a parameter for the ZPE. From the substantially different derivation presented in Ref. [1] and reviewed in Section 3.1, it is evident that the Meyer-Miller Hamiltonian eq 1 is only a special case of the comprehensive mapping Hamiltonian eq 20 when $\Gamma_{nm} = \gamma \delta_{nm}$. Parameter $\gamma$ of eq 1 in fact represents a parameter for the diagonal element of a diagonal commutator matrix. That is, commutator matrix $\Gamma$ is equal to the product of a constant ($\gamma$) and an identity matrix. It hints that parameter $\gamma$ of eq 1 can be negative. Figure 3 demonstrates that the negative value for parameter $\gamma$ is indeed possible as well as useful for the spin-boson model. The most important feature of Figure 3 is that parameter $\gamma$ of eq 1 in principle should not to be interpreted as a conventional ZPE parameter. This is confirmed by more results for the spin-boson model at even zero temperature in Ref. [3]. Figure 3 demonstrates that eCMM with $\gamma \in (-1/F, 1/2]$ is overall superior to either Ehrenfest dynamics or FSSH, especially in the low-temperature, strong coupling, or nonadiabatic regions.

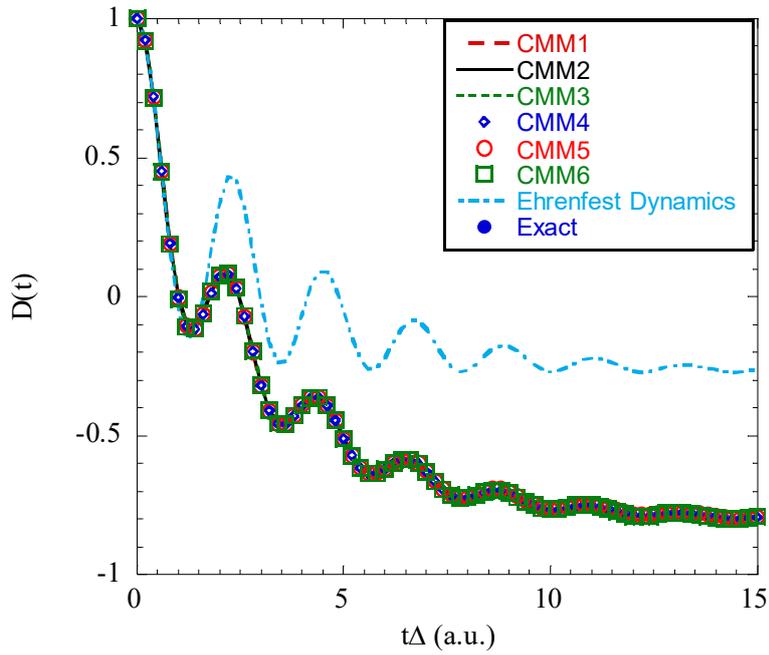

**Figure 2** (Color). Population difference $D(t) = P_{1\leftarrow 1}(t) - P_{2\leftarrow 1}(t)$ of the spin-boson Hamiltonian with the Debye bath at finite temperature. The six mapping Hamiltonian models of Ref. [1] are used in CMM1-CMM6 (i.e., for eCMM approaches with $\gamma = 0$), respectively. Ehrenfest dynamics results as well as exact results are demonstrated for comparison. (Reproduced with permission from Ref. [3]. Copyright 2019 American Institute of Physics.)

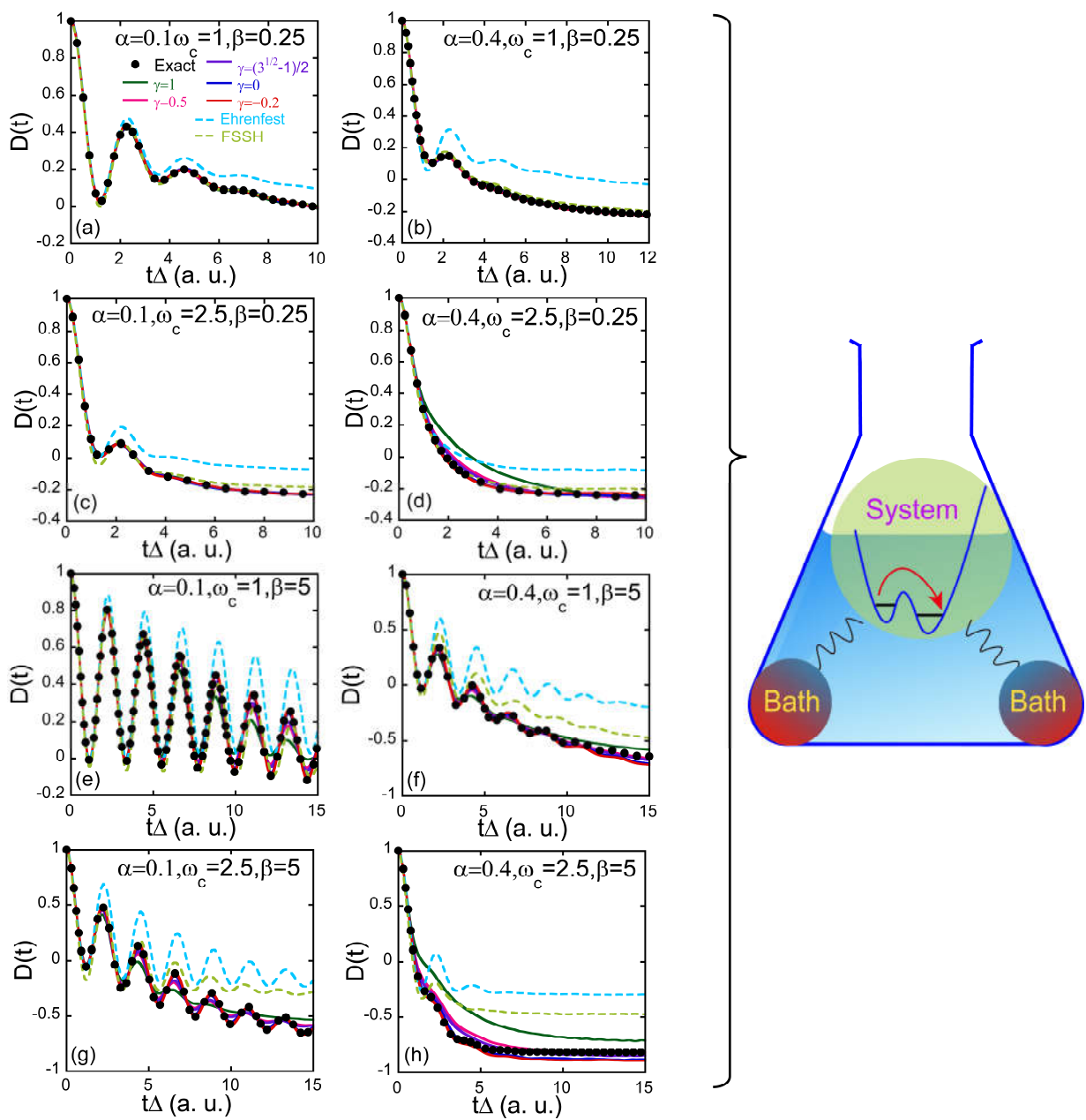

**Figure 3.** (Color). Population difference $D(t) = P_{1\leftarrow 1}(t) - P_{2\leftarrow 1}(t)$ of the spin-boson Hamiltonian with the Ohmic bath at finite temperature. Ehrenfest dynamics and FSSH are compared to the eCMM approach with $\gamma = 1$, $0.5$, $(\sqrt{3}-1)/2$, $0$, and $-0.2$. See Ref. [4] for more details. (Adapted with permission from Ref. [4]. Copyright 2021 American Chemical Society.)

### 4.2. Three-Electronic-State Photodissociation Models

The second set of benchmark models are the coupled three electronic-states where the PESs are Morse oscillators as proposed in Ref. [53]. The PESs and coupling terms are depicted in Refs. [42, 53]. The set of models mimic ultrafast photo-dissociation processes in molecular systems. Since they involve relatively local coupling terms, the state-state coupling is nearly zero at short times. It is expected that the Born-Oppenheimer limit is indicated in short-time dynamics of these models. Figure 4 demonstrates that the eCMMcv results are in good agreement with the exact data. The performance of eCMMcv is superior to that of eCMM in all three models. While eCMMcv yields more accurate results than eCMM, eCMMcv is also less sensitive to the value of parameter $\gamma$. This is mainly because the Born-Oppenheimer limit is satisfied in eCMMcv. Figure 4 indicates that eCMMcv performs better than Ehrenfest dynamics and FSSH for the three ultrafast photo-dissociation models.

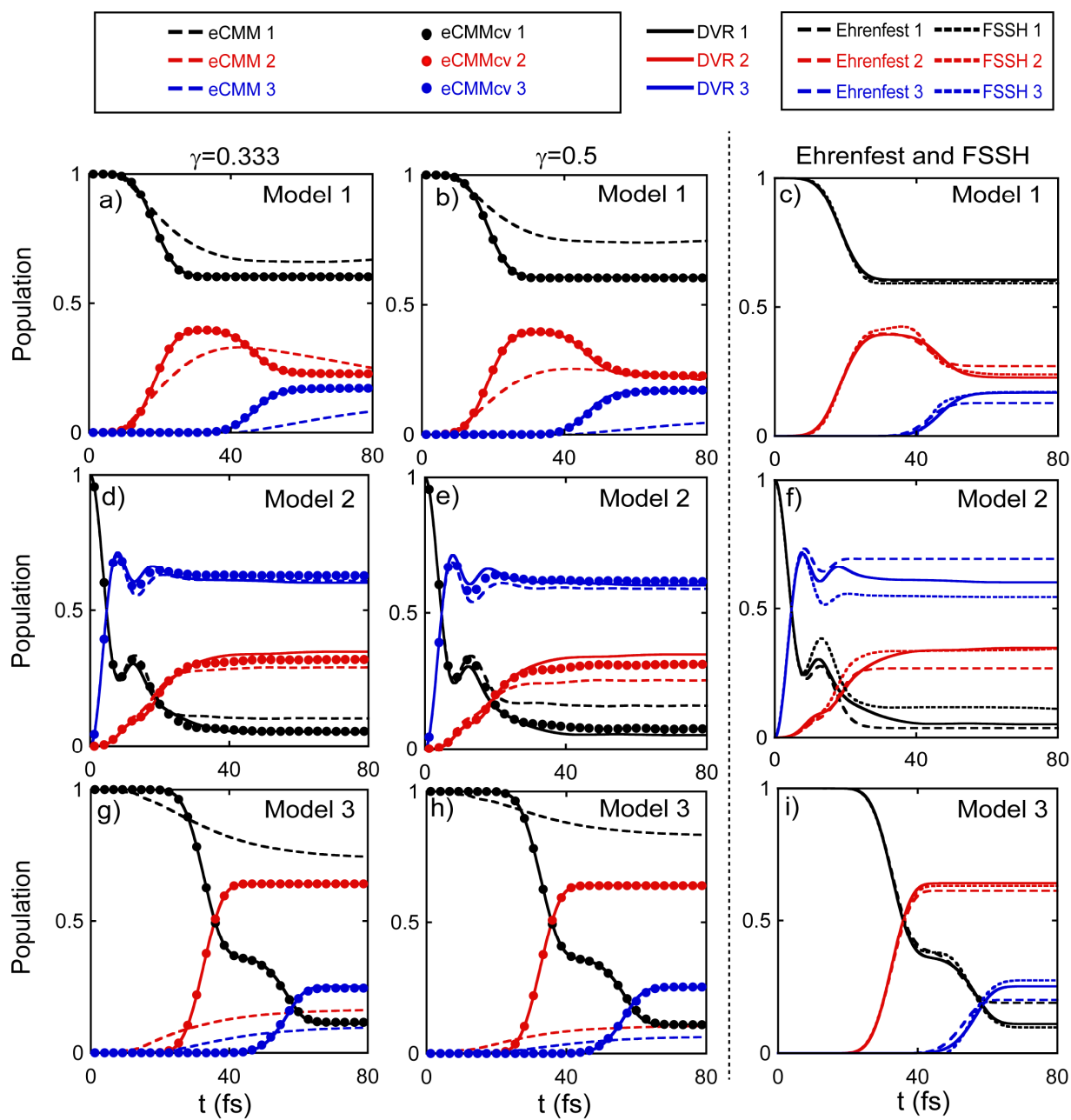

**Figure 4:** Population dynamics of the three electronic states for photo-dissociation models[53]. The eCMM as well as eCMMcv results are obtained with $\gamma = 0.333$ or $\gamma = 0.5$. Parameter $\gamma$ is recommended in the region, $\left[\left(\sqrt{F+1}-1\right)/F,\, 1/2\right]$. Black, Red and Blue markers represent Populations on state 1, 2 and 3, respectively. See Ref. [42] for more details. (Adapted with permission from Ref. [42]. Copyright 2021 American Chemical Society.)

## 4.3  Seven-Site Fenna-Matthews-Olson (FMO) Monomer

The third application is the site-exciton model of the Fenna-Matthews-Olson (FMO) monomer of green sulfur bacteria[54]. The FMO monomer involves seven photosynthetic pigments (Bacteriochlorophyll). When each pigment is denoted by a site or state, a seven-site model can be established. The parameters of such a typical model are described in Ref. [55]. Fifty effective modes (nuclear DOFs) per site are enough for converging the simulation results. We study both diagonal and off-diagonal elements of the reduced density matrix for the sites (i.e., electronic DOFs) of the photosynthetic system. While the diagonal element represents the population of the site (shown in Figure 5), the off-diagonal elements are the electronic coherence terms (demonstrated in Figure 6). Panel 5(e) or Panel 6(e) implies that Ehrenfest dynamics performs poorly for the FMO monomer. In contrast, either eCMM or eCMMcv is capable of producing much more reasonably good results when parameter $\gamma \in \left[ \left( \sqrt{F+1} - 1 \right) / F, \, 1/2 \right]$. Figures 5-6 show that the overall performance of eCMMcv is slightly better than eCMM for the FMO monomer model. Figure 7 depicts population dynamics of the FMO monomer at three different temperatures. It is shown that the relaxation time scale increases as the temperature decreases. It is often demanding for "numerically exact" methods to study the zero temperature behavior of such as the FMO monomer model. The eCMM/eCMMcv approaches predict that the time scale of the oscillation of the population (of site 1) at 0K lasts significantly longer than that at 77K.

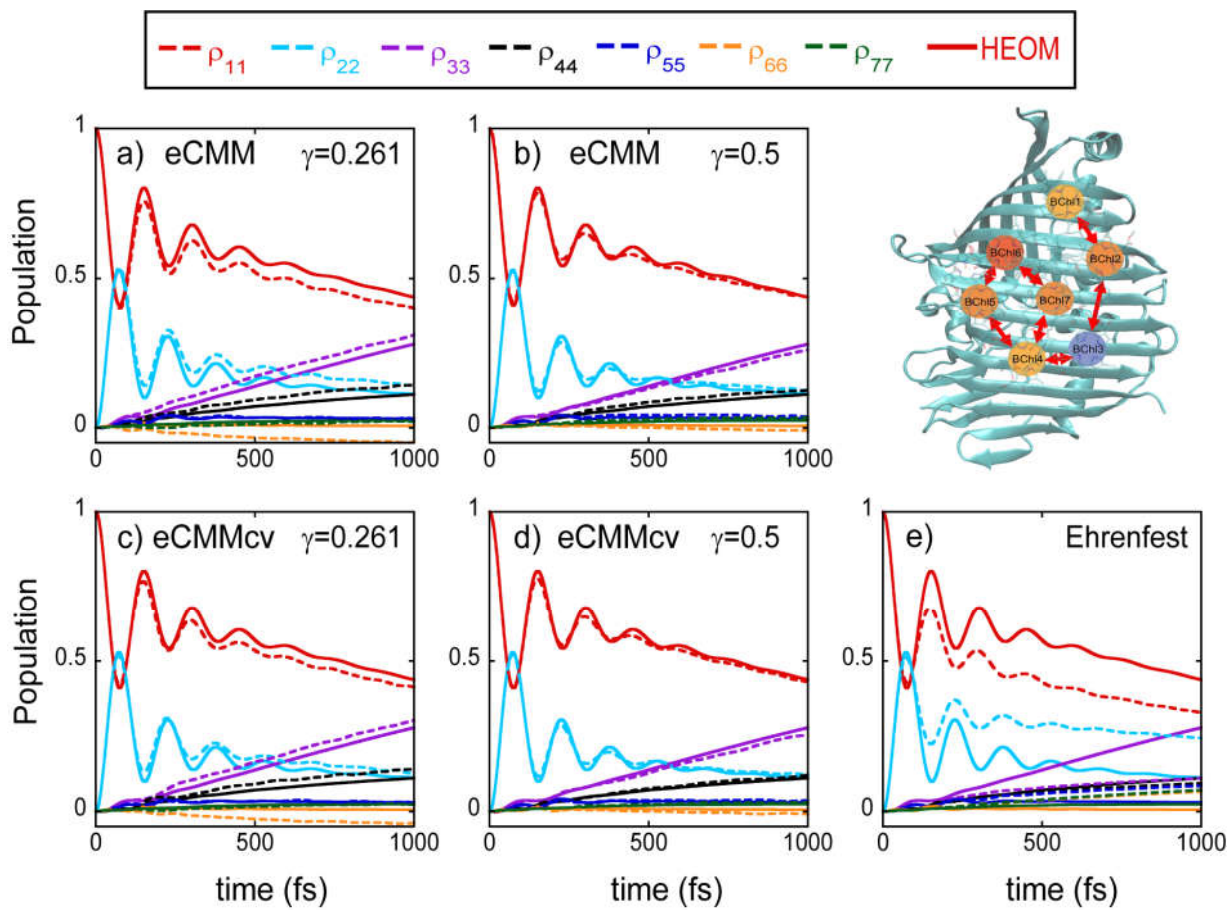

**Figure 5:** Population dynamics of the seven-site FMO monomer model at 77K, where the first pigment (Site 1) is initially excited. Solid lines represent exact results by HEOM. See Ref. [42] for more details. (Adapted with permission from Ref. [42]. Copyright 2021 American Chemical Society.)

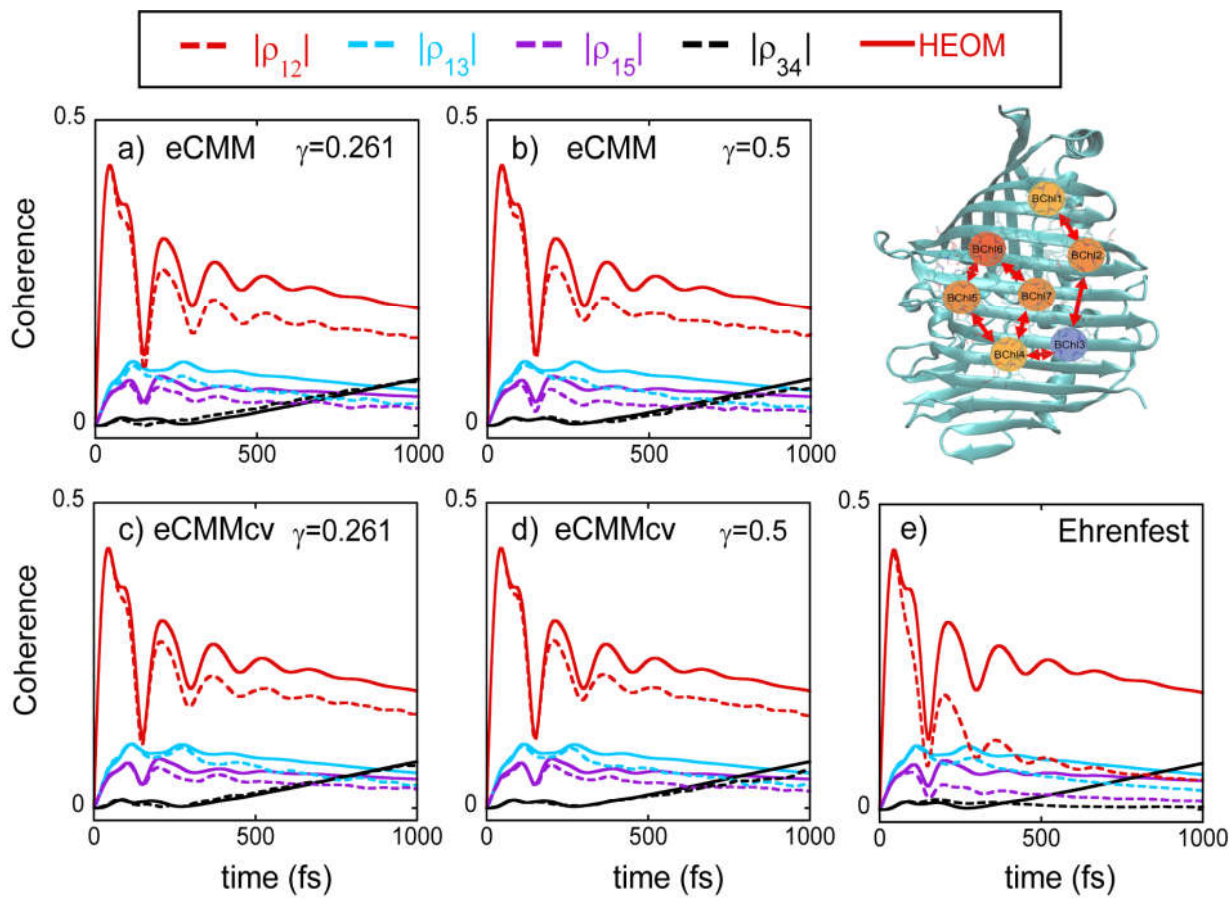

**Figure 6:** Same as Figure 5, but for the absolute values of the electronic coherence terms. See Ref. [42] for more details. (Adapted with permission from Ref. [42]. Copyright 2021 American Chemical Society.)

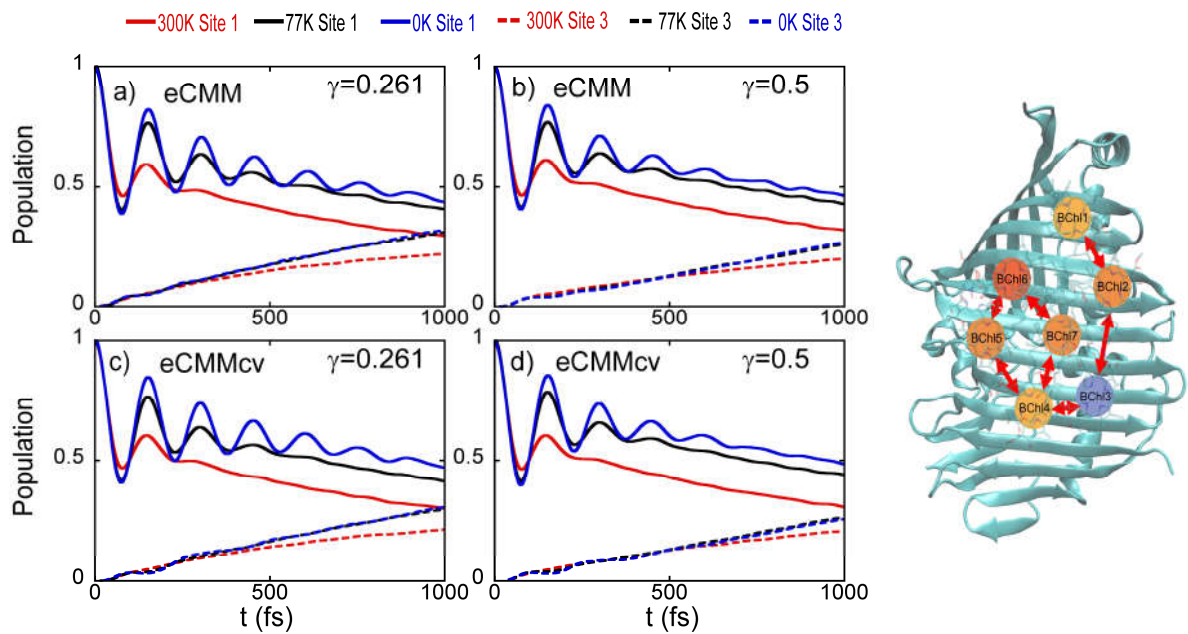

**Figure 7**. Population dynamics of the seven-site FMO monomer model at different temperatures. See Ref. [42] for more details. (Adapted with permission from Ref. [42]. Copyright 2021 American Chemical Society.)

### 4.4  Atom-in-Cavity Models

In recent progress on laser techniques and optical microcavities[56], the interaction between the external (intense) field and the molecular can induce nonadiabatic couplings between different electronic states and then alter chemical dynamics in a qualitative level. The fourth application of eCMM/eCMMcv is a set of benchmark model systems that describe an atom with a fixed position in a one-dimmensional lossless cavity. The coupling between different atomic electronic states is created by the interaction between the transition (dipole) moment and the optical cavity field.[38, 57-59]. Four hundred field modes are used to obtain converged results. More details on the atom-in-cavity models and on implementation of eCMM/eCMMcv are available in Ref. [42].

As demonstrated in Figure 8 and Figure 9, the results that Ehrenfest dynamics or FSSH predicts significantly deviate from exact cavity-modified chemical dynamics even at very short times. In contrast, both eCMM and eCMMcv achieve considerably better performance. Either of eCMM and eCMMcv is capable of semi-quantitatively capturing the negative (positive) spike in the population of the ground (excited) electronic state around $t=1800$ a.u., which corresponds to the reabsorption and re-emission process of the earlier emitted photon by the atom.

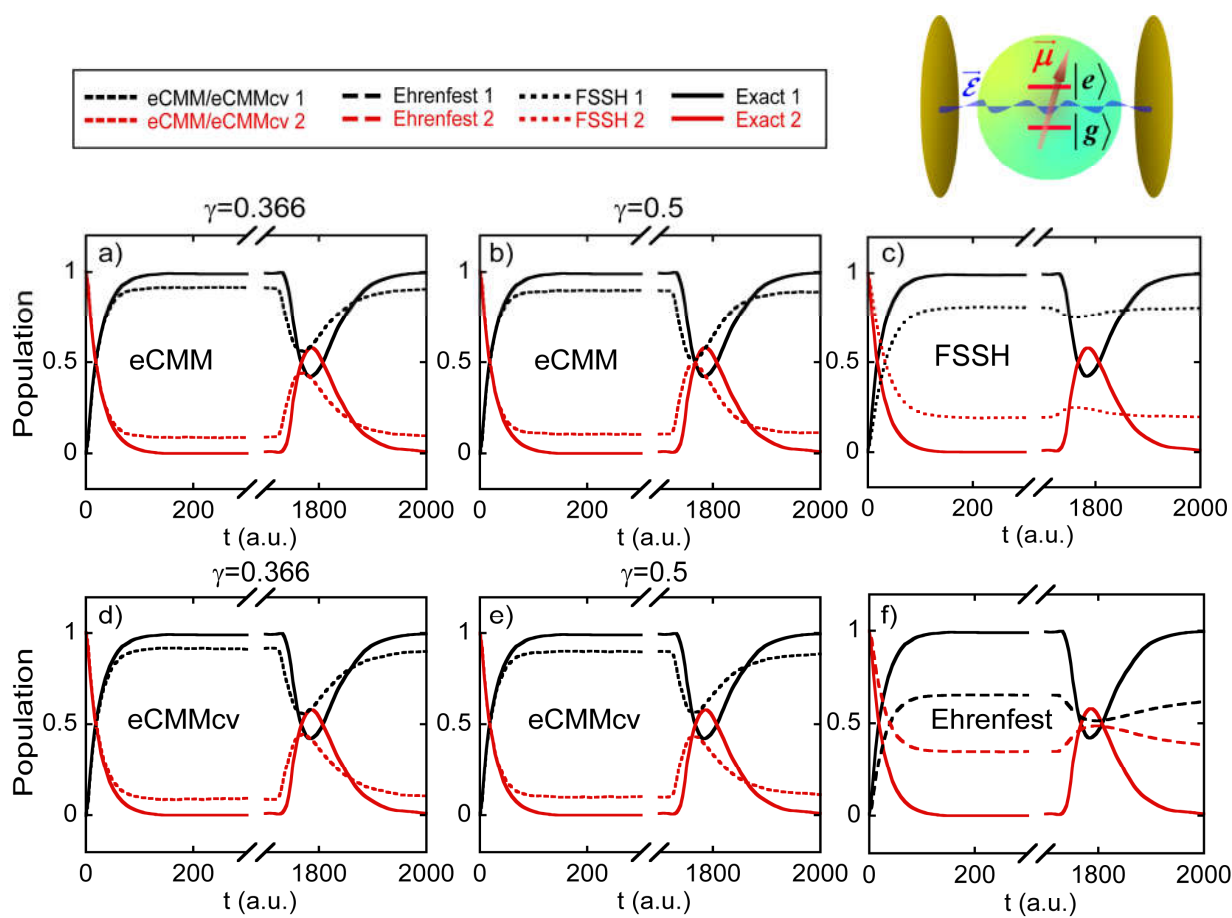

**Figure 8.** Population dynamics for the two-electronic-state model for an atom in optical cavity. The highest atomic electronic excited state is initially actived. Black and Red colors represent Population of State 1 and that of State 2, respectively. See Ref. [42] for more details. (Adapted with permission from Ref. [42]. Copyright 2021 American Chemical Society.)

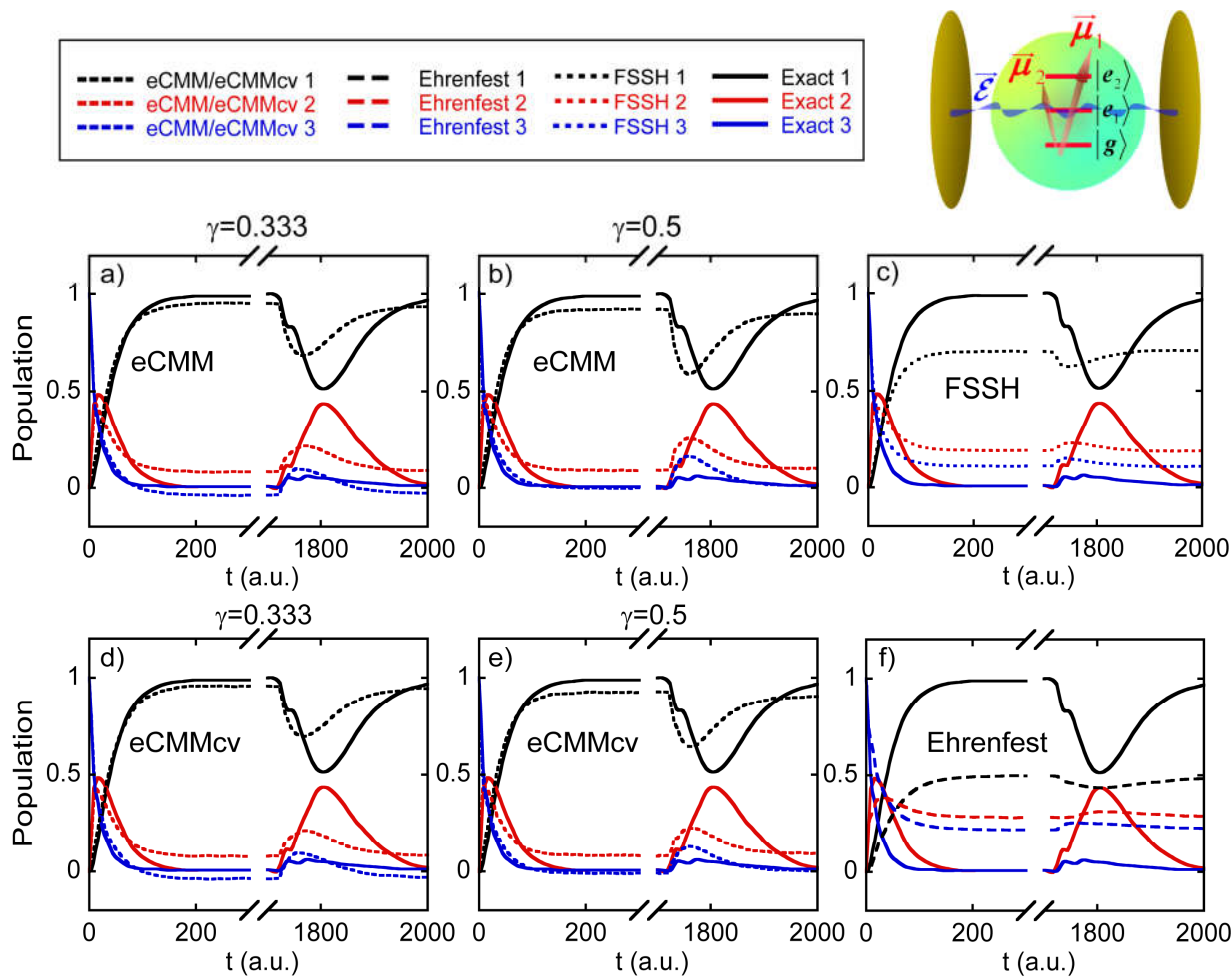

**Figure 9.** Same as Figure 8, but for the three-electronic-state model for an atom in optical cavity. Black, Red, and Blue colors represent Population of State 1, that of State 2, and that of State 3, respectively. See Ref. [42] for more details. (Adapted with permission from Ref. [42]. Copyright 2021 American Chemical Society.)

## 5. Relation between the coupled multi-state Hamiltonian and the second-quantized many-electron Hamiltonian

It is shown in Ref. [2] that an isomorphism exists between the coupled multi-state Hamiltonian and the second-quantized many-electron Hamiltonian with only 1-electron interactions.

Equation 21, model I of Ref. [1] is reminiscent of the Cartesian classical second-quantized many-electron Hamiltoian model of Refs. [60, 61] which has been used to study electron transport phenomena. Any phase space mapping approaches for nonadiabatic dynamics in the preceding discussion can be useful for investigating nonequilibrium dynamics in electron transport processes.

As a proof-of-concept example, we study the conventional resonant level (Landauer) model[62, 63] for a quantum dot state coupled to two electrodes. The parameters of the resonant level (Landauer) model are also presented in Ref. [2]. Figure 10(a) demonstrates that any phase space mapping Hamiltonian models included in the unified framework of Ref. [1] are capable of reproducing exact results for the Landauer model. Figure 10(b) then shows that the steady current of the Landauer model shows an "S"-shape voltage dependence that is more distinct as the temperature decreases (or $\beta$ increases). A crossover behavior exists in the relation between the steady current and the temperature for the Landauer model[2]. The isomorphism indicates the phase space approach with the comprehensive mapping Hamiltonian (such as eq 20 or eq 28) will in principle also be useful for studying the second-quantized many-electron Hamiltonian system where both electronic and nuclear motions are involved in experimentally related electron transport processes.

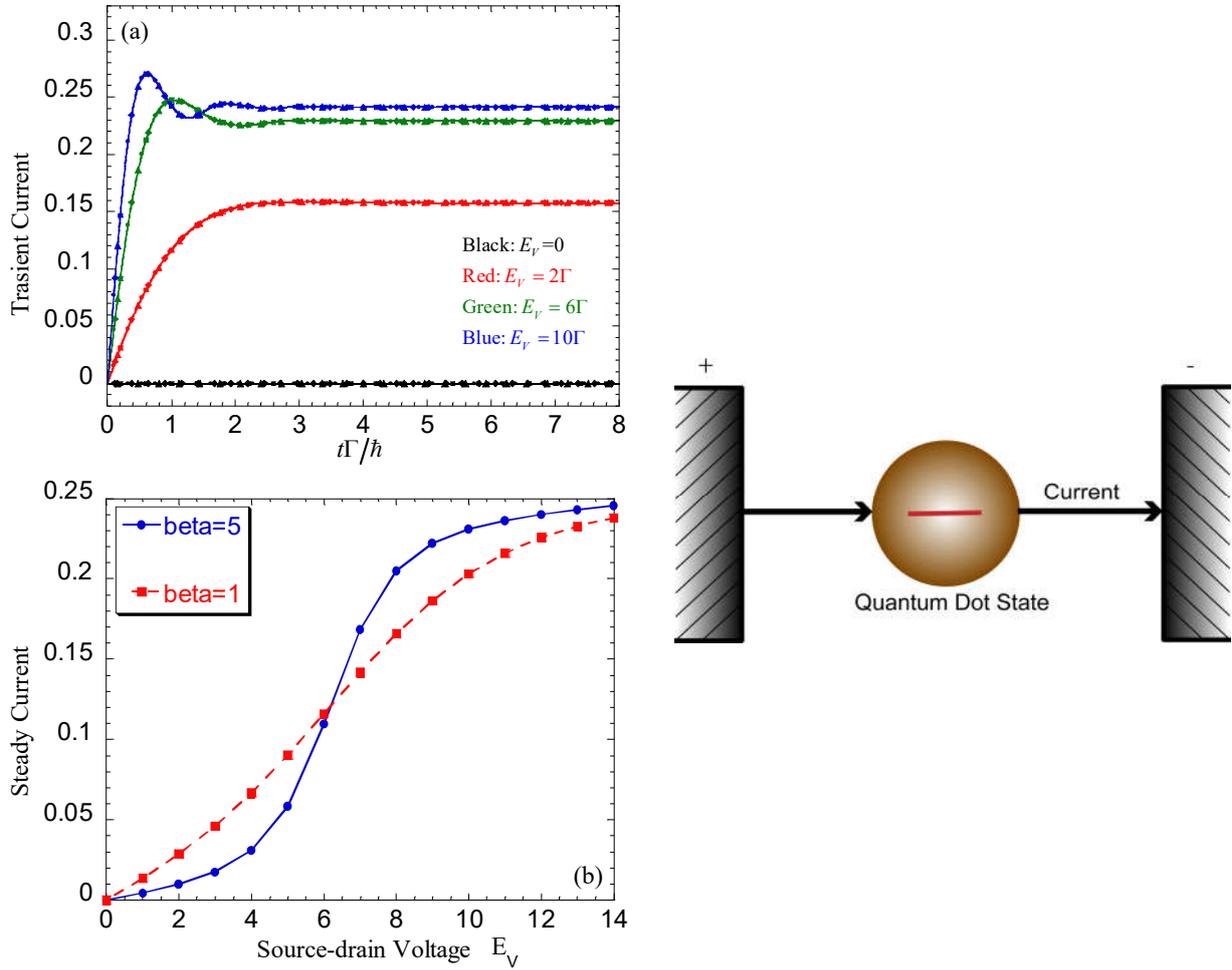

**Figure 10** (Color). (a) Transient currents of the Landauer model with different source-drain voltages $E_V$. (b) The value of the steady current as a function of $E_V$ for $\beta = 5$ and that for $\beta = 1$. ($\Gamma = 1, \hbar = 1, e = 1$) (Panels a and b are adapted with permission from Ref. [2]. Copyright 2017 American Institute of Physics.)

## 6.  CONCLUDING REMARKS

We have developed a general exact phase space mapping formulation for nonadiabatic molecular systems that include continuous nuclear DOFs as well as a finite set of electronic states. It sets the scene for constructing new nonadiabatic dynamics methods. Although it is easy to express the exact EOMs (e.g., quantum Liouville equation) in the general phase space mapping formulation, it is extremely difficult, if not impossible, to obtain exact dynamics when nuclear DOFs are included in realistic molecular systems. Trajectory-based dynamics yielded by the comprehensive mapping Hamiltonian leads to exact results for the electronic DOFs in the frozen nuclei limit, but becomes an approximation to nonadiabatic quantum dynamics where nuclear and electronic EOMs are coupled.

It is evident from various (condensed phase) applications that such trajectory-based phase space mapping dynamics is capable of significantly outperforming Ehrenfest dynamics[10] as well as FSSH[11], leading to reasonably accurate results for the evolution of the population as well as the electronic coherent term. Parameter $\gamma \in \left[ \left( \sqrt{F+1} - 1 \right)/F, \, 1/2 \right]$ is recommended in the exact mapping kernel (eq 43). The more general and comprehensive mapping Hamiltonian (e.g., eq 20) evidently outperforms the conventional Meyer-Miller Hamiltonian (eq 1), as shown in Figures 4-9 on the comparison between eCMM and eCMMcv. (This conclusion holds when we consider the most recent version of SQC dynamics with triangle window functions[30], as shown in the Supporting Information of Ref. [42].) We expect that the comprehensive mapping Hamiltonian with the commutator matrix will be useful for on-the-fly nonadiabatic dynamics[39, 40]. The expression of the phase space mapping approach in the adiabatic representation is available in the Supporting Information of Ref. [42]. Provided that the initial condition is fixed, results of quantum nonadiabatic dynamics are independent of the representation of the electronic basis. The important criterion is satisfied in the phase space mapping approach of this Account.

Regardless of successful applications of the trajectory-based approximate nonadiabatic quantum dynamics approaches in the phase space formulation, a major drawback is that the detail balance[64, 65] for *both electronic state and nuclear DOFs* is not rigorously satisfied when the entire system reaches thermal equilibrium. To the best of our knowledge, no approximate practical nonadiabatic dynamics methods have been developed for fundamentally addressing this challenge for realistic systems where nuclear quantum effects should be considered. It is shown in our recent work that the sign problem is inevitable in path-based or trajectory-based approaches in obtaining the thermal equilibrium distribution of general nonadiabatic systems[66]. Although the strategies of Refs. [67-69] as well as semiclassical dynamics of Ref. [23] are interesting, it remains a challenge to practically employ them for systematically improving over the mapping Hamiltonian dynamics for complex (large) nonadiabatic systems.

Since the celebrated Meyer-Miller mapping Hamiltonian model[13] was proposed, phase space mapping theory has been still growing to meet the needs of nonadiabatic quantum dynamics of complex (large) systems. This Account highlights our recent progress: a unified framework for constructing the phase space mapping Hamiltonian[1], a general phase space formulation of quantum mechanics for nonadiabatic systems where the electronic DOFs are mapped onto constraint space and the nuclear DOFs are mapped onto infinite space[3, 4, 42], and an isomorphism between the mapping phase space approach for nonadiabatic systems and that for nonequilibrium electron transport processes[2]. Given the considerable interest in photoexcited or external field modified dynamic processes, it is expected many developments and extensions will be underway.

## ■ AUTHOR INFORMATION

**Corresponding Author**

*E-mail: jianliupku@pku.edu.cn

**ORCID**

Jian Liu: 0000-0002-2906-5858

Xin He: 0000-0002-5189-7204

Baihua Wu: 0000-0002-1256-6859

**Notes**

The authors declare no competing financial interest.

**Biographies**

**Jian Liu** received his BS from the University of Science & Technology of China in 2000 and PhD from the University of Illinois at Urbana-Champaign in 2005. He then spent a postdoc at the University of California, Berkeley and a research associate at Stanford University before he joined Peking University as an associate professor in 2012. His research interests have been focused on the development of trajectory-based methods for studying quantum statistics and dynamics of complex (large) molecular systems.

**Xin He** is a third year PhD student in theoretical chemistry at Peking University. He earned his BS from Peking University in 2019. His research interests are in nonadiabatic quantum dynamics.

**Baihua Wu** is a third year PhD student in theoretical chemistry at Peking University. He received his BS from the University of Science and Technology Beijing in 2019. His research interests are in nonadiabatic chemistry.

■ **ACKNOWLEDGMENT**

This work was supported by the National Natural Science Foundation of China (NSFC) Grant No. 21961142017, and by the Ministry of Science and Technology of China (MOST) Grant No. 2017YFA0204901. We acknowledge the High-performance Computing Platform of Peking University, Beijing PARATERA Tech CO., Ltd., and Guangzhou supercomputer center for providing computational resources. We acknowledge Zhihao Gong for his contribution to Refs [4,42]. We also thank William H. Miller, Wenjian Liu, Yi Zhao, Jianshu Cao, Eli Pollak, Oleg V. Prezhdo, Weihai Fang, and Jiushu Shao for useful discussions.

■ **REFERENCES**

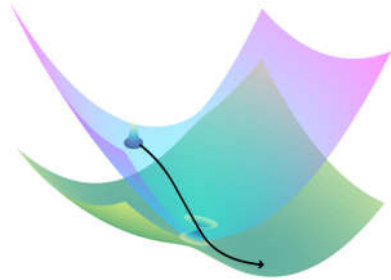

$$[\hat{x}^{(n)}, \hat{p}^{(m)}] + [\hat{x}^{(m)}, \hat{p}^{(n)}]$$
$$= 2i(\Gamma_{nm} + \Gamma_{mn})$$